\documentstyle[aps,preprint,pre]{revtex}
\tightenlines
\begin{document}
\input epsf
\draft

\title{Geography in a Scale-Free Network Model}
\author{C. P. Warren$^{1}$\cite{cemail}, L. M. Sander$^{1}$\cite{lemail}, 
and I.M. Sokolov$^{2}$\cite{iemail}}

\address{$^1$Michigan Center for Theoretical Physics, Department of Physics, University of Michigan, Ann Arbor, Michigan, 48109-1120\\
$^2$Institut f\"ur Physik, Humboldt-Universit\"at zu Berlin, Invalidenstr. 110, D-10115 Berlin, Germany.}
\date{\today}
\maketitle

\begin{abstract} 
We offer an example of an network model with a power law degree distribution, 
$P(k)\sim k^{-\alpha}$, for nodes but which nevertheless has a well-defined
geography and a nonzero threshold percolation probability for $\alpha>2$, the 
range of real-world contact networks.  This is 
different from the $p_c=0$ for $\alpha<3$ results for the original well-mixed 
scale-free networks.  In our \textit{lattice-based scale-free 
network}, individuals link to nearby neighbors on a lattice.  Even 
considerable additional small-world links do not change our conclusion of nonzero 
thresholds.  When applied to disease propagation, these results suggest that random 
immunization may be more successful in controlling human epidemics than previously 
suggested if there is geographical clustering.
\end{abstract} 


\section{Introduction}

The idea that a scale-free network has a zero percolation threshold
\cite{HavlinAlpha3,Newman}, $p_c=0$, has sparked a good deal of interest 
lately\cite{LloydMay,VespignaniEp,LloydMaySIR,VespignaniImm}.  In this paper, we offer 
an example of a scale-free network (in the sense of a power law degree distribution) that 
has very different properties than the original Barabasi-Albert (BA) network.  We find that 
replacing the preferential attachment and well-mixed structure of the BA network with 
two-dimensional (2d) clustering makes the percolation threshold nonzero.  Further, 
percolation on our network can be mapped onto the problem of SIR epidemic propagation
\cite{Grassberger,We}.  Thus, these percolation results have practical implications 
for the control of real world epidemics.  Specifically, in contrast to recent claims
\cite{VespignaniImm,Barapc0}, a random immunization program might successfully 
eradicate an epidemic on a scale-free network, such as a sexually transmitted disease.  
Furthermore, we will show that the addition of a sizable number of 
random links, or small worlds bonds, to the lattice does not change the result of a nonzero 
percolation threshold in our network.  

In our network model, nodes are embedded on a 2d lattice.  They asymptotically
have a power law distribution of degree $k$, $P(k)\sim k^{-\alpha}$, but they connect 
to \textit{the $k$ nearest neighbors on the lattice}, not randomly chosen nodes.  The 
relevance of this to disease propagation is that for epidemics such as sexually
transmitted diseases (STDs) and many other human diseases it is known that 
the number of contacts of different individuals differs over a wide range.  The 
distributions of the numbers of sexual contacts have been traced in 
several studies \cite{Sex} and are known to show fat tails which approximate
power-law behavior: The number of sexual partners of an individual $k$
during a year is distributed according a power law with $2<\alpha <3$\cite{Liljeros}. 
These distributions possess a mean value, $\overline{k}<\infty$, but they lack a dispersion 
and thus show large, universal fluctuations.

To model such behavior, Pastor-Satorras and Vespignani as well as Lloyd and May 
\cite{VespignaniEp,LloydMaySIR} have proposed a scale-free network with preferential attachment 
to already highly connected nodes, or hubs. 
In the original model due to Barabasi and Albert\cite{Barabasi}, the network grows 
sequentially: Each newly introduced node brings $m$ proper bonds; each of them is 
attached to one of the $n$ nodes already existing; and the probability of 
attachment is proportional to the number $k_{i}$ of the already existing bonds 
of this node. This model leads naturally to a power-law probability distribution of 
the degree of the node (number of its bonds), $P(k)\propto k^{-3}$. Infection 
propagation on a scale-free network was considered in Pastor-Satorras and 
Vespignani \cite{VespignaniEp,VespignaniImm}.  They find that a 
scale-free network is very robust against the random removal, or immunization, of 
the nodes. A giant component still persists even if almost all nodes in the system are 
eliminated, i.e. the critical threshold $p_c=0$ for all $\alpha\leq3$.  

Asserting that the network of human sexual contacts has the scale-free construction 
and structure of the BA model would have important consequences for controlling the 
epidemics of sexually transmitted diseases: The immunization of the large part of the 
population would be a useless measure.  Instead, one would have to concentrate on the most 
active agents, which may be hard to identify.

Although it may be a reasonable model for growing technological networks, such
as the World Wide Web, where physical distance is not an issue, the
scale-free construction appears to us to be unnatural as a model for the disease
transmission.  Its ``well-mixed'' spanning character and the
absence of any underlying metric (i.e. the impossibility of definition of a
geographical ``neighborhood'') are unrealistic.
(In simple terms, although some people often use airplanes, a large portion 
of the Earth's population does not.)  Sexual contacts may have well-mixed 
properties \textit{locally} within a neighborhood, town or city, but in general 
not globally.  For nearly the entire population, physical distance at some scale 
matters.  There is no reason that the topology (a structure of 
connections) of a network of sexual contacts is the same as that of the World
Wide Web.

Eguiliz and Klemm \cite{Klemm} recently offered a different example of a 
scale-free network with nonzero SIS epidemic thresholds for $\alpha=3$.  In
our model, nonzero thresholds are present for any $\alpha>2$, so it addresses
the region of interest.  Also, in their model,
nodes with a particular degree, or number of links, are more likely to mix 
with those of a unlike degree.  In their otherwise well-mixed network, 
highly-connected nodes are more likely to connect with sparsely-connected nodes,
and vice versa.  This is a disassortative, or degree-anticorrelated, network.  
However, Newman\cite{NewAssort} argues that social interaction networks are
assortative(i.e. degree-correlated) as discussed later.  In our network, just like the 
BA network, there is \textit{no} degree correlation.  Instead, the 
nonzero threshold comes from the underlying local 2d clustering structure that 
would seem naturally present in real epidemics, structure that is not present 
in Eguiluz and Klemm's model. Rosenfeld, Cohen, ben-Avraham and Havlin\cite{Havlin} 
have a method of embedding a scale-free network into a 2d lattice that is similar 
to ours but not identical, and they explore the dimensional properties of their 
network. 

\section{Model: Bond Percolation with Variable Number of Links}

The basic version of our model on a lattice preserves the local geometrical 
properties of a 2d lattice.  For that reason, we call it a \textit{
lattice-based scale-free network}. Our model is a variant of an
old two-dimensional circle model of continuum percolation (a
lattice-based ``inverse Swiss cheese'' model\cite{Feng} 
with variable radius of ``holes'').  Starting from the sites on the lattice 
we assign each node its number $k_{i}$ of proper bonds defining its 
``radius of action'' $R$ and connect it to all the nodes within the radius 
$R$. The distribution of the radii is taken so that the number of proper bonds,
the bonds put down within $R$, follows the $k^{-\alpha }$-law for 
large $k$ and cuts off at the nearest  neighbor distance.  In $d$ 
dimensions, if the number of proper bonds of a site is to be 
distributed according to $P(k)\propto k^{-\alpha }$, it follows
that $\Pi(R)\simeq R^{-\beta }$ with $R\geq1$
and $\beta=d(\alpha -1)+1$.  Nodes having a bond in common are
considered connected. The bond $ij$ connecting nodes $i$ and $j$ is
counted only once, whether it belongs to the set of the proper bonds of node 
$i$, of node $j$, or both.  Note that in our model $\alpha$ and $\beta$
are parameters; they do not arise naturally as in the BA network.

First, we show that the mean number of bonds per node in our model has
the same asymptotic power-law behavior as the distribution of the number of the
proper bonds if $\alpha>2$.  We calculate the mean number of bonds connecting a
node $i$ to nodes $j$ outside $R_i$. This is also given by the mean
number of the nodes at distance $r>R_i$ from the given one for which $R_j>r$. 
The probability to find $R_j$ larger than $r$ is 
$\int_{r}^{\infty }\Pi(r^{\prime })dr^{\prime }$.  Thus, 
the mean number of nodes connected to $i$ from outside is 
\[
M_{R_{i}}= \int_{R_{i}}^{\infty}drS_{d}r^{d-1}\int_{r}^{\infty }\Pi(R_{j})dR_{j}.
\]
where $S_{d}$ is the surface of the $d$-dimensional unit sphere ($S_{1}=1$, 
$S_{2}=2\pi $, etc.). In general, taking $\Pi(R_{j})\simeq 
R_{j}^{-d\alpha +d-1}$,
\[
M_{R_{i}}\simeq R_{i}^{-d(\alpha -2)}
\]
which tends to zero for large $R_{i}$, as long as the mean number of bonds per node 
exists(i.e. $\alpha >2$). Note that the number $M_{R}=\sum_{n=1}^{\infty }nq_{n}$ 
(where $q_{n}$ is the probability that a node considered has exactly $n$
bonds outside of $R$) is larger than the probability $Q_{R}=\sum_{n=1}^{\infty
}q_{n}$ to have at least one bond coming from outside $R$, and thus for $%
\alpha >2$, $Q_R$ tends to zero as $R$ grows. This means that
the probability distribution of the number of the ``proper'' bonds and the
actual number of the bonds of a node follow the same asymptotic pattern for $%
k$ large. The mean number of bonds per node is given by $\overline{k}%
=\int_{0}^{\infty }(S_{d}R^{d-1}+M(R))\Pi(R)dR$ and converges for $\alpha >2$.

Now, consider bond percolation on this network.  Beginning with a 2d square grid of $N$ bare 
nodes, randomly add $Np$ disks (i.e. choose radii $R_i$ from $\Pi(R_i)$ and fill up the proper bonds
to the nodes that don't yet have them).  For what $\alpha$ will there be a
nonzero percolation threshold, $p_c>0$?

We note that percolation on our network 
differs from conventional lattice percolation or continuum percolation in that for any 
finite lattice there is a possibility of a node having a radius of action so large that 
it spans the entire $L$ x $L$ lattice.  The probability of 
drawing such a giant disk scales as $P_{giant}=Prob(r\geq L)\sim L^{1-\beta}$.  The average 
number of disks put down before such a giant disk is encountered is 
approximately $1/P_{giant}$. In $d$ dimensions, that number corresponds to $p^{*}L^d$ 
disks, where $p^{*}$ is the average threshold for adding a giant disk on an $L$ x $L$ lattice.  
Thus, $1/P_{giant}=p^{*}L^d$ and solving for the average threshold $p^{*}$ for an $L$ x $L$ lattice, 
we obtain
\begin{equation} 
	p^{*}\sim L^{\beta-1-d}=L^{d(\alpha-2)}, 
\label{onespan} 
\end{equation} 
so that $p^{*}\rightarrow 0$ as $L\rightarrow\infty$ for $\alpha <2$.
Since adding a giant disk is only one of several ways of spanning the lattice,
$p^{*}$ is an upper bound for the percolation probability $p_c$.  Thus, $p_c(L)\rightarrow 0$ as 
$L\rightarrow\infty$ for $\alpha <2$ as well.

The general SIR (Susceptible-Infected-Recovered) model of the infection propagation on
a simple lattice can be mapped on to the percolation problem on that same lattice
\cite{Grassberger,We,NewmanPerc}.  The bonds present in the percolation problem
correspond to successful propagation of the disease from an infected individual to 
a susceptible.  Thus, a subcritical cluster in the percolation problem corresponds
to a subcritical epidemic, an epidemic that dies out.  Infection propagation is possible 
(i.e. a giant component of a graph exists) if a finite fraction of the individuals are
infected. 

Nodes without disks might be thought of as ``immune'', but the analogy is not complete.  Since we 
are doing bond percolation, nodes are always present.  ``Immune'' nodes will lack
the proper bonds, but they can still have outside bonds, which would not be the case for
a truly immune individual.  Thus, our model will actually \textit{underestimate} the true 
epidemic threshold.  Consider the difference in the simple example of two
disks shown in Fig. 1.  In our model, just the presence of those
two disks would mean that the epidemic spans.  Thus, for this simple example, $p_c=2/64=1/32$.  
In the proper epidemic model, one would need an additional susceptible node in the overlap 
of the two disks for the epidemic to be able to span the lattice.  Thus, $p_c>1/32$.  The 
correspondence of our disk model to an epidemic on a scale-free network is not perfect but 
adequate for our purposes.  

\section{Simulation}

We found percolation thresholds using the Newman-Ziff algorithm
\cite{newman-ziff}.  Disks were randomly added to an $L$ x $L$ lattice at different
sites until a cluster spanned the lattice.  Figure 2 shows the average percolation 
threshold $p_c(L)$ as a function of $1/L$.  It appears from the plot that 
there is a finite percolation threshold for not just $\alpha>3$ but $2<\alpha<3$ as well, 
the region of interest for real world epidemics.  Thus, if these real world epidemic 
are not well mixed but rather dominated by local geometry, they will have a finite 
threshold.  For $\alpha<2$, the results are consistent with $p_c=0$ and seem to scale
according to Eq. (\ref{onespan}) for sufficiently large $L$.  The transition is gradual.

In comparing simulations, we find that $p^{*}(L)$ is significantly larger than $p_c(L)$, even for
$\alpha=1.6$.  However, as one can see from Fig. 2, the slopes of $p_c(L)$ and $p^{*}(L)$ as a 
function of lattice size match fairly well for $\alpha\leq1.9$.

\section{Small World Links}
Consider now the same model with small world links, bonds that connect two randomly chosen
nodes, added as well.  In the context of disease propagation, this is included to model
infrequent, distant contacts -- occasional airline travel, as it were.  Suppose $Np$ disks 
are present along with $Np\phi$ random links as well. 
How do these additional links affect the percolation threshold?  On a lattice with small 
world links, the simple spanning criterion of conventional lattice percolation is not 
appropriate.  Random links may possibly connect two sites on opposite boundaries at a low 
concentration with no infections between, but this is hardly captures the idea of a 
sustained epidemic.  As we argued previously\cite{We}, the proper criterion for the 
percolation threshold is when the fraction of the lattice occupied by the largest cluster 
$M(p)$ becomes a finite fraction of the $L$ x $L$ lattice as $L \rightarrow \infty$.  
We use finite size scaling on $M(p,L)$ to find this value as in \cite{We}.  Fig. 3 shows 
simulation results for the 
percolation threshold $p_{SW}(\phi)$ for the disk lattice with small world bonds as a 
function of the fraction of added random links $\phi$.  Clearly, for $\alpha>2$, the 
addition of even a sizable number of small world links does not result in $p_c=0$.

One approach to finding the percolation threshold on a small world lattice is to ignore 
the random links and consider the subcritical clusters of the lattice as nodes 
on a random network\cite{We,NewmanPerc}.  The random links become the links for this random
network.  Using this approach, the percolation threshold for a random network with $A$ 
nodes and $B$ bonds is $A=2B$.  If $\bar{n}$ is the average subcritical cluster size, there 
will be $A = N/\bar{n}$ nodes on such a network and 
$B=Np\phi$ bonds.  The subcritical percolation clusters will scale as 
$\bar{n}\propto |p-p_c|^{-\gamma}$\cite{Stauffer}, 
where $p_c$ is the threshold without random links and $\gamma$ is a characteristic exponent.  
We can estimate a new threshold $p_{SW}(\phi)$ for the small world lattice implicitly
from the relation 
\begin{equation} 
	\phi=\frac{K|p_{SW}-p_c|^{\gamma}}{p_{SW}}, 
\label{SWeq}
\end{equation} 
where $K$ is a nonuniversal constant.  What is $\gamma$?  For conventional 2d lattice 
percolation, $\gamma=43/18\approx2.39$.  As shown in Figure 3, eq. (\ref{SWeq}) with the 
2d value for $\gamma$ work quite well for $\alpha\geq3$.  The $\phi=0.316$ data point 
was used to calculate $K$.  

For smaller $\alpha$, specifically the measured 
exponent of $\alpha=2.5$ for females in the Swedish sexual network\cite{Liljeros}, this 
approximation breaks down.  One possibility for this breakdown is an incorrect value 
of $\gamma$.  Perhaps the presence of more large disks changes $\gamma$.  
Using the $\phi=0.0316$ and $\phi=0.1$ data points, a value of $\gamma=1.94$ was 
calculated, but the fit is still inadequate.  Some crossover seems to be occurring 
which we do not understand.  We believe the most likely explanation is that 
implicit in the nodes-on-a-random-graph approximation is the assumption that the 
subcritical islands are equally likely to be chosen by the random links.  However, if 
there are several large subcritical islands on the lattice, this assumption 
apparently breaks down.  These large islands gain more attachments but not enough to 
reduce the percolation threshold to zero.

\section{Discussion}
The primary reason for the difference between the BA model and our model is the distance 
and number of connections between the highly connected hubs.  With the preferential 
attachment of the BA model, nodes are more likely to attach to hubs and particularly 
connect two hubs, offering numerous \textit{very short} network paths between hubs and 
thus to a sizable portion of the population.  These numerous pathways between make 
percolation more likely.  

In contrast, the local clustering of geography in our model lengthens the network pathways 
between hubs.  A disease on our network would be much easier to control.  Similarly, as also 
mentioned by Newman\cite{NewAssort} in an degree anticorrelated network, highly connected nodes 
are more likely to be connected to sparsely connected nodes, thus lengthening the network 
distance between the hubs.  A moderate number of random links in our model will not change 
these results even though these links are effectively preferentially attached.  This is because 
the bulk of the attachments are made through local clustering not preferential attachment.  At 
any rate, in order to characterize the percolation properties of a scale-free network, one 
needs to know more than the degree distribution and the degree correlation distribution.

Newman\cite{NewAssort} notes that social networks tend to 
be assortative or degree-correlated, and he concludes that because of this they may not be 
conducive to immunization efforts.  However, the networks that he cites are career-related 
collaborations such as movies and coauthorships, which may not reflect the nature of the 
network of physical interaction that would be relevant to disease propagation.  In addition, 
with 2d local clustering we have provided an alternative reason that immunization 
efforts may indeed be fruitful as in the case of other highly infectious diseases such as 
polio and smallpox.  

CPW is supported by a Rackham Interdisciplinary and Collaborative Research Grant.  Special thanks
to the Center for the Study of Complex Systems for computer resources.

\begin{figure}
\epsfxsize=\hsize \centerline{\epsfbox{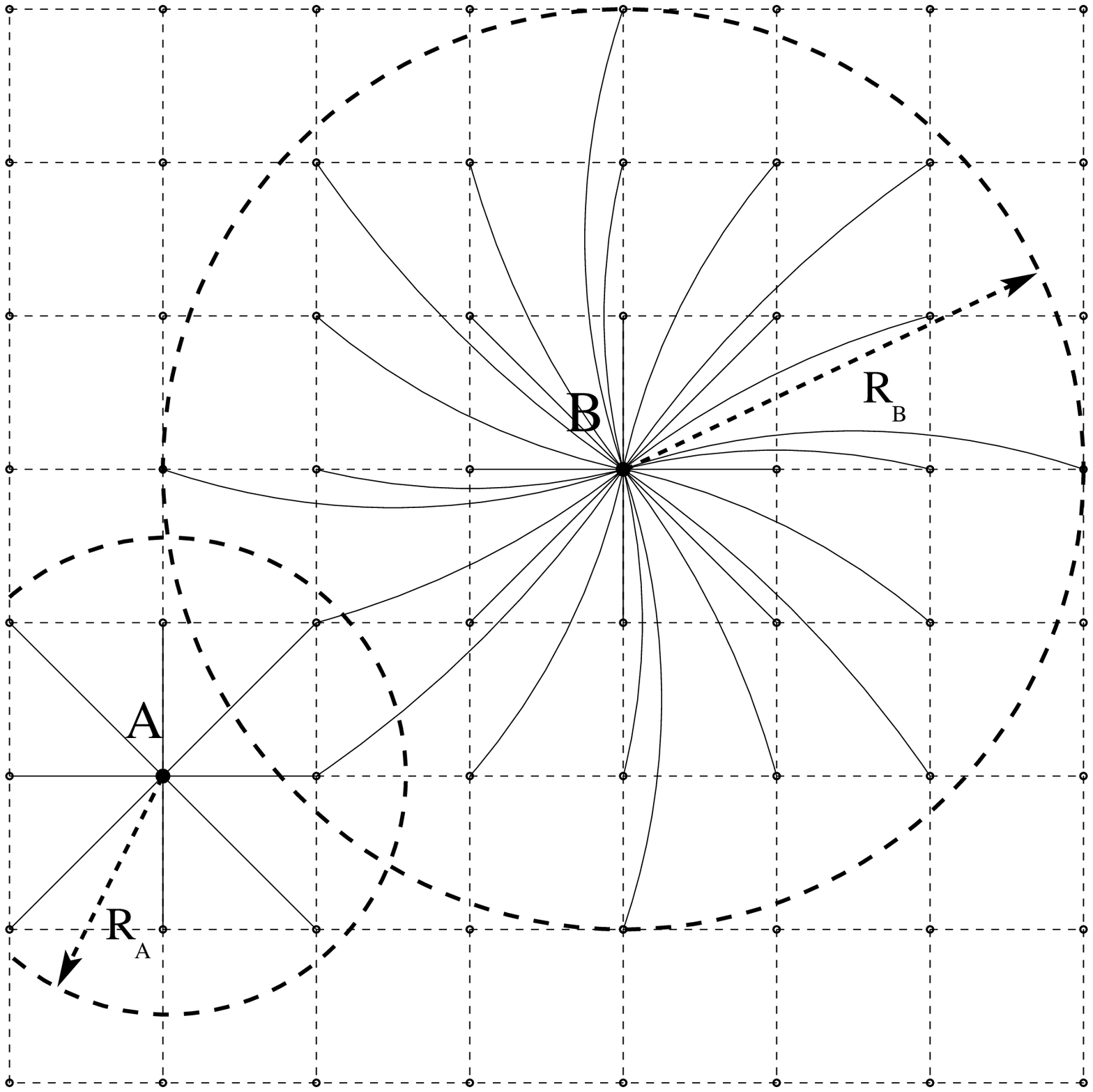}}
\caption{
An example of the basic model.  One disk with radius $R_A$ is centered around node $A$,
and another with radius $R_B$ is centered about $B$.  All nodes within each disk are attached to
the central node.  The bonds of these two disks span the $8$ x $8$ lattice from left to right.
}\label{f1}  
\end{figure}

\begin{figure}
\epsfxsize=\hsize \centerline{\epsfbox{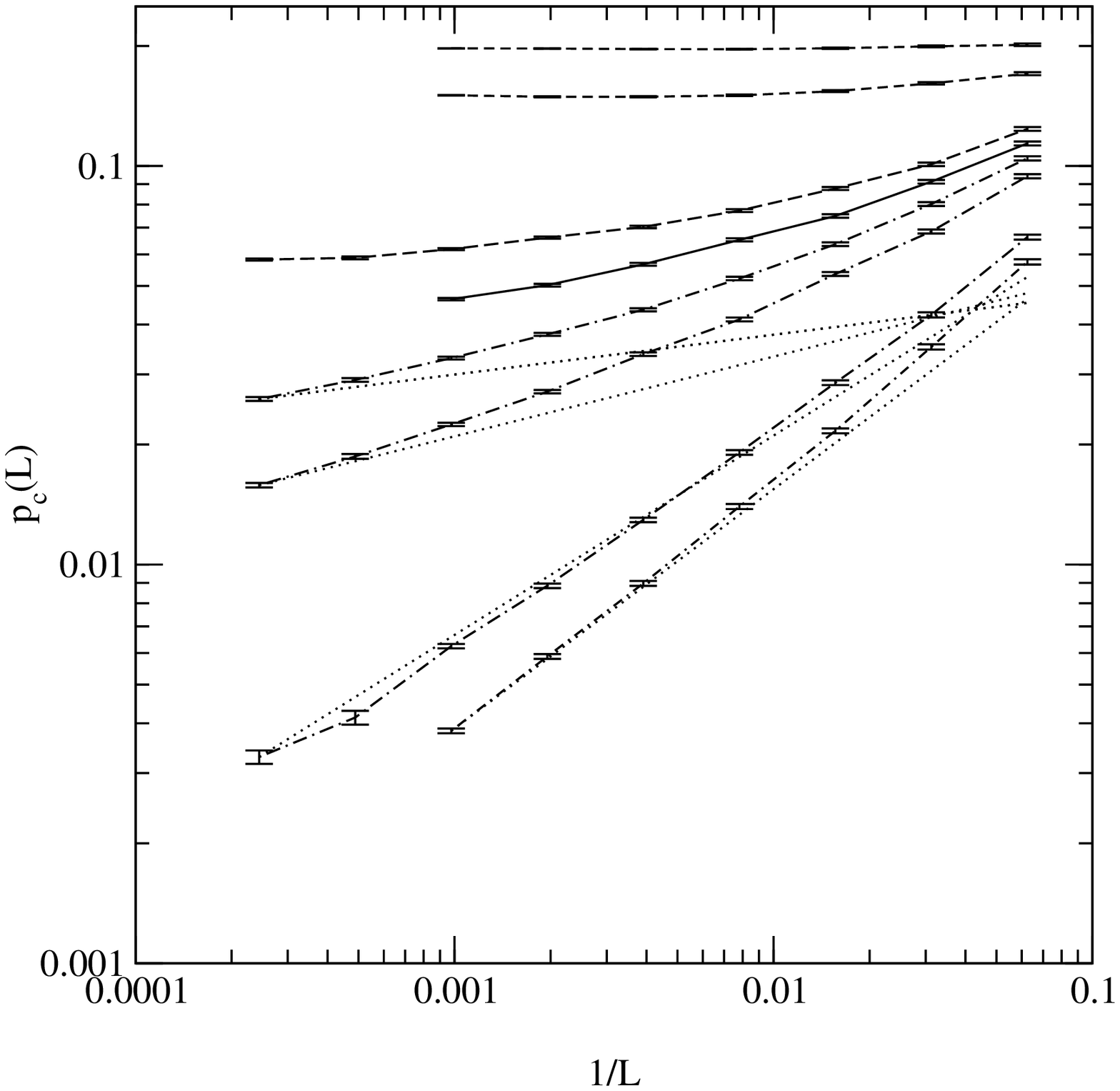}}
\caption{
Percolation threshold on an $L$ x $L$ lattice as a function of  $1/L$ for various powers $\alpha$ 
in the lattice-based scale-free model showing an apparent change in behavior at $\alpha=2$.  From 
top to bottom, the dashed lines are $\alpha=2.5$, $2.3$, and $2.05$, the solid line is $\alpha=2$, 
and dash-dot lines are $\alpha=1.95$, $1.9$, $1.75$, and $1.7$.  2000 simulations were run for 
$L\leq1024$ data points, and 1000 were run for $L>1024$ data points.  The dotted lines are the
\textit{slopes} of $p^*$ for the presence of giant spanning disks.  These lines are fitted either
to $L=1024$ or $L=2048$, and they are not $p^*$, which is significantly larger.   
}\label{f2}  
\end{figure}

\begin{figure}
\epsfxsize=\hsize \centerline{\epsfbox{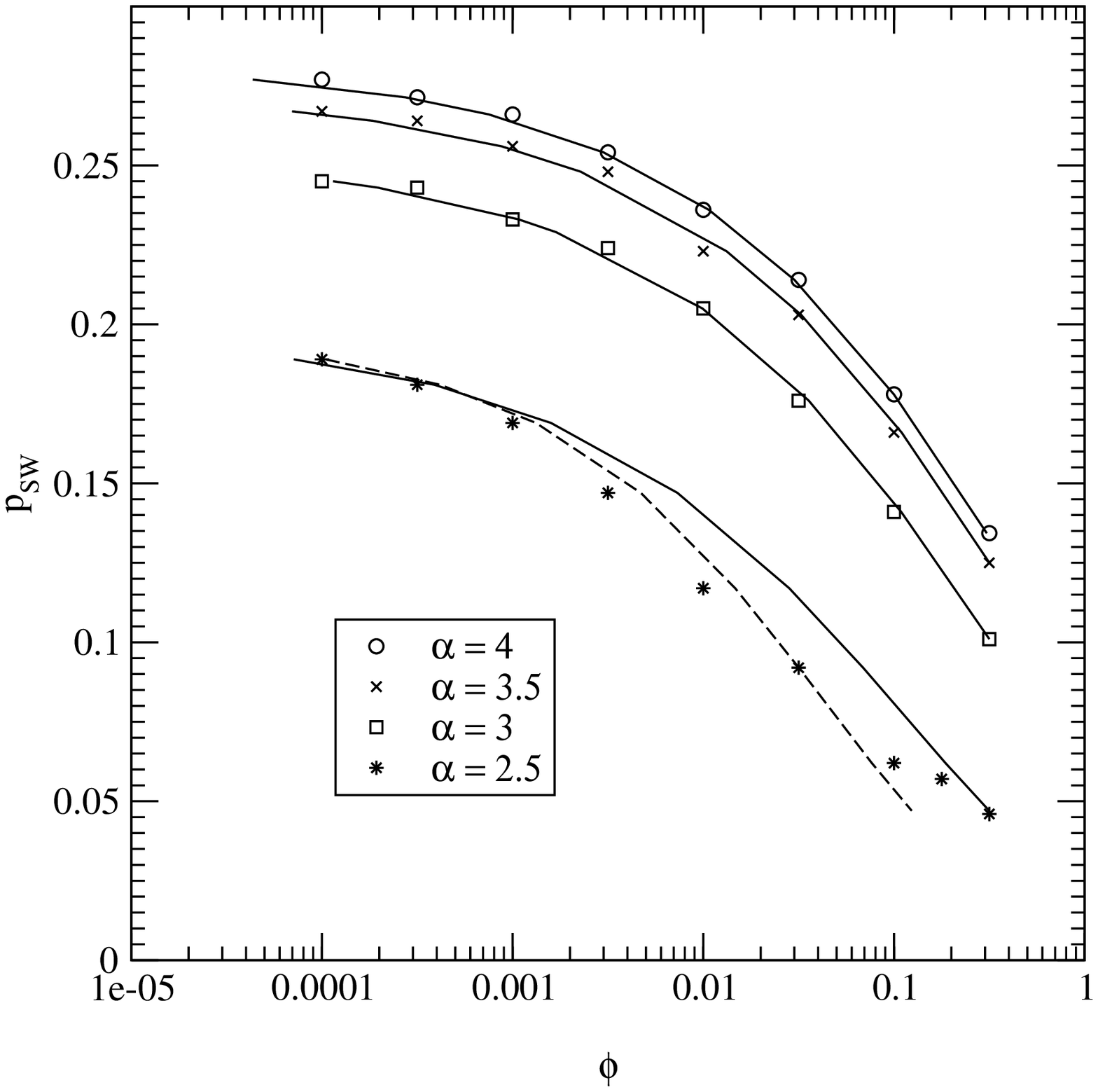}}
\caption{
Average percolation threshold as $Np\phi$ small world links were added to the lattice-based 
scale free model.  These infinite lattice thresholds were extrapolated from finite lattice 
simulations.  Solid lines are predictions from the nodes-on-a-random-graph estimate with 
the giant cluster exponent $\gamma = \gamma_{2d}\approx2.39$.  Each line is fitted to the 
$\phi=0.316$ value.  The dashed line uses a calculated $\gamma=1.94$ value and is fitted 
to the $\phi=0.1$ value.
}\label{f3}  
\end{figure}

\end{document}